\documentclass[12pt]{article}

\usepackage{epsf,amsmath}
\usepackage[dvips]{graphicx}

\begin{document}

\begin{center}

PROPOSAL OF POLARIZED $^3$He$^{++}$ ION SOURCE FOR JINR ACCELERATOR
COMPLEX

\

 V.V. Fimushkin$^{\dag}$, E.D. Donets, I.V. Gapienko, Yu.A. Plis,
 Yu.V.~Prokofichev, V.P.~Vadeev

\

Joint Institute for Nuclear Research,  Dubna\\
$\dag$ {\it E-mail: fimushkin@jinr.ru}

\end{center}

\vskip 5mm

\begin{center}
\begin{minipage}{150mm}
\centerline{\bf Abstract} It is proposed to develop a source of
polarized $^3$He$^{++}$ ions for the JINR Accelerator Complex  on
the basis of its polarized deuteron source by feeding its
radio-frequency dissociator with $^3$He gas to produce metastable
atoms, using an existing sextupole magnet and adding a low-field rf
transition units.

The radio-frequency transitions of the atomic states of helium-3 in
the metastable state are studied. The Schroedinger equations in the
uncoupled basis $|\psi_e,\psi_h>$ and also in the basis of
stationary states are received. Results of computer simulations
agree with published data. The possibility to use two types of the
weak field transitions in the helium-3 with different frequencies to
get positive or negative values of the helion polarization is shown.

 Ionization and accumulation of the polarized helions may be
carried out by the electron beam ion source (EBIS) in the reflex
mode of operation. The project ionizer parameters are the following:
electron energy 10~keV, effective current 5~A, ion trap length 1~m,
$^3$He$^{++}$ ion beam intensity   $\approx2\times10^{11}$~ions per
fast extraction pulse of 8~$\mu$s.
\end{minipage}
\end{center}

\vskip 10mm

\section{Introduction}

The goal is to make a source of  polarized $^3$He$^{++}$ ions
(helions) on a basis of the polarized deuteron source for
NUCLOTRON-M. The rf dissociator is fed with helium-3 gas for
production of $^3$He atoms in the metastable 2$^3$S$_1$ state.
Stern--Gerlach separation with a sextupole magnet and  rf
transitions in a weak magnetic field are used for nuclear
polarization of the metastable atoms. Ionization to $^3$He$^{++}$
and accumulation of the polarized helions will be carried out by the
electron beam ion source (EBIS) in the reflex mode of operation
\cite{EDonets}.

We note that magnetic moments of a helion  and neutron are close in
the value:\\ $\mu_h=-2.127\mu_N$ and $\mu_n=-1.913\mu_N$, where
$\mu_N$ is the nuclear magneton.

Earlier, the Laval University  group (Canada) \cite{Slobodrian}
polarized $^3$He atoms in the metastable state 2$^3$S$_1$ (a
lifetime of 7860 s) with the electron angular momentum $J=1$ and
then ionized them to $^3$He$^+$ in an electron impact ionizer.

The cold cathode discharge source  of metastable atoms produced a
flux of $6\times10^{15}$ atoms/s sterad with an average velocity of
$2.5\times10^5$~cm/s. The ionization potential of  metastable atoms
is quite low, 4.6 eV, compared to 24.6 eV for atoms in the ground
state. The ionizer operated  in a mode that discriminated between
metastable and ground-state atoms, thus producing $^3$He$^+$ from
the nuclear polarized metastable atoms.

The subsequent ionization to $^3$He$^{++}$ was effected by stripping
in the base of the Van de Graaf accelerator at 7.5~MV.

The angular momentum of a $^3$He($2^3$S$_1$) atom is the vector sum
$\vec{F}=\vec{J}+\vec{I}$, where $\vec{J}$ and $\vec{I}$ correspond
to the electronic ($J=1$) and nuclear ($I=1/2$) angular momenta,
respectively.

The spin-dependent part of the  Hamiltonian for $^3$He atoms in the
metastable state $2^3$S$_1$ is
\begin{equation}
\hat{H}=-\mu_J\vec{J}\vec{B}(t)-\mu_h\vec{\sigma_h}\vec{B}(t)-\frac
13\Delta W\vec{\sigma_h}\vec{J}, \label{eq19}
\end{equation}
where $\vec{\sigma_h}$ is the Pauli spin matrices of the helion,
$\vec{J}$ is the electron spin matrices ($J=1$), $\vec{B}(t)$ is the
magnetic field strength,  $\Delta W$ is  the hyperfine splitting of
the $^3$He atom in the  state $2^3$S$_1$;\\ $\Delta W=
4.4645\times10^{-24}$ J$=\hbar\times
4.2335\times10^{10}$ rad/s,\\
$\mu_J=2\mu_e=-1.85695275\times10^{-23}$ J/T$=-\hbar\times
1.76085977 \times10^{11}$ rad/s T,\\ $\mu_h=-1.07455\times10^{-26}$
J/T$=-\hbar\times 1.0189 \times10^8$ rad/s~T.

Consider at first the case of static magnetic field $B$ directed
along the z-axis. The six hyperfine structure stationary states and
corresponding eigenvalues are  obtained by diagonalizing the
$6\times6$ static Hamiltonian matrix  in the uncoupled $|m_h,m_J>$
basis. The wave functions of the hyperfine states at magnetic fields
$B$ and $B\rightarrow\infty$ are\\
$$
\Psi_1=\Psi(1/2, +1/2)=-\sin\beta\,\psi_h^+\psi_J^0
+\cos\beta\,\psi_h^-\psi_J^+\; \Rightarrow \psi_h^-\psi_J^+,
$$
$$
\Psi_2=\Psi(3/2, +3/2)=\psi_h^+\psi_J^+, $$
\begin{equation}
\Psi_3=\Psi(1/2, -1/2)=-\sin\alpha\,\psi_h^+\psi_J^-
+\cos\alpha\,\psi_h^-\psi_J^0\; \Rightarrow \psi_h^-\psi_J^0,
\label{eq20}
\end{equation}
$$
\Psi_4=\Psi(3/2, +1/2)=\cos\beta\,\psi_h^+\psi_J^0
+\sin\beta\,\psi_h^-\psi_J^+ \;\Rightarrow \psi_h^+\psi_J^0,
$$
$$
\Psi_5=\Psi(3/2, -1/2)=\cos\alpha\,\psi_h^+\psi_J^-
+\sin\alpha\,\psi_h^-\psi_J^0 \;\Rightarrow \psi_h^+\psi_J^-,
$$
$$
\Psi_6=\Psi(3/2, -3/2)=\psi_h^-\psi_J^-, $$

where $\sin\beta=\sqrt{A_+}\,,\;\cos\beta=\sqrt{1-A_+}\,;\;
\sin\alpha=\sqrt{A_-}\,,\;\cos\alpha=\sqrt{1-A_-}\,;$
$$
A_+=\frac 12(1-\frac{x+1/3}{\sqrt{1+\frac 23 x+x^2}}),\;A_-=\frac 12
(1-\frac{x-1/3}{\sqrt{1-\frac 23 x+x^2}}); $$
$$
x=\frac{B}{B_c},\;\;B_c=\frac{\Delta
W}{-\mu_J/J+\mu_I/I}=\frac{\Delta W}{-\mu_J+2\mu_h}=0.2407\, {\rm
T}. $$

\begin{figure}[h]
 \centerline{
 \includegraphics[width=50mm,height=50mm]{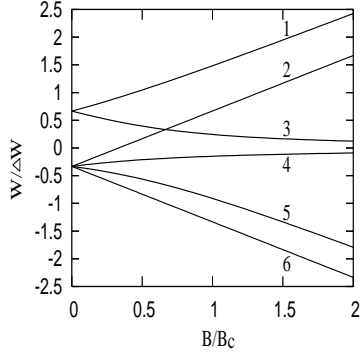}}
 \caption{Scheme of the $^3$He(2$^3$S$_1$)  states showing the  hyperfine
 structure and Zeeman splitting}
\label{Fig1}
\end{figure}

The Breit--Rabi diagram of six Zeeman hyperfine components of this
metastable state is shown in Fig. \ref{Fig1}, where the numbers
correspond to those of the wave functions $\Psi_1-\Psi_6$.

The energies of the  states $\Psi_1-\Psi_6$ are

\begin{equation*}
W_1=\frac{\Delta W}{6}-\mu_J\frac{B}{2}+\frac{\Delta
W}{2}\sqrt{1+\frac{2}{3}x+x^2},
\end{equation*}
\begin{equation*}
 W_2=-\frac{\Delta W}{3}-\mu_JB-\mu_IB,
\end{equation*}
\begin{equation*}
W_3=\frac{\Delta W}{6}+\mu_J\frac{B}{2}+\frac{\Delta
W}{2}\sqrt{1-\frac{2}{3}x+x^2},
\end{equation*}
\begin{equation*}
W_4=\frac{\Delta W}{6}-\mu_J\frac{B}{2}-\frac{\Delta
W}{2}\sqrt{1+\frac{2}{3}x+x^2},
\end{equation*}
\begin{equation*}
W_5=\frac{\Delta W}{6}+\mu_J\frac{B}{2}-\frac{\Delta
W}{2}\sqrt{1-\frac{2}{3}x+x^2},
\end{equation*}
\begin{equation*}
 W_6=-\frac{\Delta W}{3}+\mu_JB+\mu_IB,
\end{equation*}
where
\begin{equation*}
x=\frac{B}{B_c},\,\,\,B_c=\frac{\Delta W}{-\mu_J+2\mu_I}=0.2407\;
{\rm T}.
\end{equation*}

Earlier, the SATURNE group \cite{Beauvais} reported the results of
tests conducted with the use of the known HYPERION polarized ion
source fed with $^3$He gas. The dissociator was made of a Pyrex tube
with a 2.2 mm diameter nozzle cooled to 80--100 K. The gas flowed
for only 3 ms each cycle. The peak rf power was 6 kW at 19 MHz. The
ionizer with a reflex electron beam yielded mostly $^3$He$^+$ ions
with a pulsed beam current of 50~$\mu$A and pulse duration 1 ms.

The difference between the sextupole magnet "on" and "off" modes was
10~$\mu$A. This value was used to estimate the helion intensity in
the proposed helion ion source.

\section{Design parameters}

The  atomic beam source under development for the polarized deuteron
source for the NUCLOTRON-M has the following characteristics:
Dissociator is fed with pulsed rf power, peak 2 kW at 35 MHz, pulse
duration 1 ms at 4 Hz. The gas flow through the nozzle 2.2 mm in
diameter is $7.4\times10^{17}$ molecules/pulse.

Stern--Gerlach separation is effected with a permanent magnet
sextupole triplet and an electromagnetic sextupole. Pole tip
magnetic fields are from 1.66 to 1.1 T. The sextupoles focus the
atomic beam into the ionizer positioned at a distance of 120 cm from
the nozzle.

\begin{figure}[h]
 \centerline{
 \includegraphics[width=50mm,height=50mm]{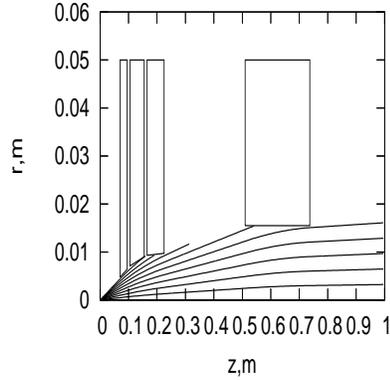}}
 \caption{ Trajectories of the atoms in the state $F=3/2,\;
 m_F=+3/2$, $v=2500$ m/s}
 \label{v=2500}
\end{figure}

\begin{figure}[h]
 \centerline{
 \includegraphics[width=50mm,height=50mm]{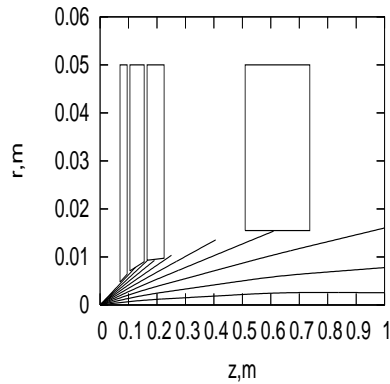}}
 \caption{ Trajectories of the atoms in the state $F=3/2,\;
 m_F=+3/2$, $v=1200$ m/s}
 \label{F=+3/2}
\end{figure}

At Figs. \ref{v=2500}, \ref{F=+3/2} the results of computer
simulation are shown for two values of the atom velocity 2500 and
1200 m/s. One may see that the $^3$He atomic beam cannot be focused
in the given configuration at the velocity $v=2.5\times10^3$ m/s.
The cooling of the atomic beam is necessary.

\section{Radio-frequency transitions}

For hydrogen the task the rf transitions of atomic spin states was
considered by E.P.  Antishev and A.S. Belov \cite{Antishev}. S. Oh
\cite{Oh}  published detailed results for weak field transitions in
deuterium. Here we solve this problem for weak  field transitions
(WFT) in $^3$He in the metastable state 2$^3$S$_1$.

 In the uncoupled $|m_h,m_J>$ state basis
$$
\Psi(t)=C_1(t)\psi_h^+\psi_J^+ +C_2(t)\psi_h^+\psi_J^0
+C_3(t)\psi_h^+\psi_J^-$$$$ +C_4(t)\psi_h^-\psi_J^+
+C_5(t)\psi_h^-\psi_J^0 +C_6(t)\psi_h^-\psi_J^-
$$
and we obtain the following  equations for the amplitudes:
$$
\frac{dC_1}{dt}=-i/\hbar \{C_1[-(\mu_h+\mu_J)B_z+\Delta W/3]
-C_2\mu_JB_x/\sqrt{2}-C_4\mu_hB_x\}
$$
$$
\frac{dC_2}{dt}=-i/\hbar \{-C_1\mu_JB_x/\sqrt{2}-C_2\mu_hB_z-
C_3\mu_JB_x/\sqrt{2}+C_4\sqrt{2}\Delta W/3-C_5\mu_hB_x\}
$$
$$
\frac{dC_3}{dt}=-i/\hbar \{\frac{-C_2\mu_JB_x}{\sqrt{2}}+
C_3[(-\mu_h+\mu_J)B_z-\Delta W/3]+C_5\sqrt{2}\Delta W/3-
C_6\mu_hB_x\} $$
$$
\frac{dC_4}{dt}=-i/\hbar \{-C_1\mu_hB_x + \frac{C_2\sqrt{2}\Delta
W}{3} +C_4[(\mu_h-\mu_J)B_z-\Delta W/3]-C_5\mu_JB_x/\sqrt{2}\}
$$
$$
\frac{dC_5}{dt}=-i/\hbar \{-C_2\mu_hB_x+C_3\sqrt{2}\Delta W/3
-C_4\mu_JB_x/\sqrt{2}+C_5\mu_hB_z-C_6\mu_JB_x/\sqrt{2}\}
$$
$$
\frac{dC_6}{dt}=-i/\hbar \{-C_3\mu_hB_x -C_5\mu_JB_x/\sqrt{2}
+C_6[(\mu_h+\mu_J)B_z+\Delta W/3]\} $$

For the polarized $^3$He atomic beam, we need two weak field
transition units  that should be placed between and after the
sextupole magnets. We note that at weak magnetic fields  ($x\ll1$),
the level distance $W_1-W_3\approx-\frac 43\mu_JB$, but
$W_2-W_4=W_4-W_5=W_5-W_6\approx-\frac 23\mu_JB$.

It is different from the case of deuterium where all the distances
between the levels at  $F=3/2$ and $F=1/2$ at weak magnetic fields
are identical and  equal $\approx-2/3\mu_eB$. This fact gives the
possibility to use two types of WFT with different frequencies to
get positive or negative values of the helion polarization.

If we realize the transition $1\rightarrow3$ in the free space
between the sextupoles,  we have after the second sextupole the pure
state 2 with $F=3/2$, $m_F=3/2$, which gives after ionization
$P\approx+1$.

If we realize the transition $1\rightarrow3$ between the sextupoles
and the transition $2\rightarrow6$  after the second sextupole, we
produce the pure state 6 with $F=3/2$, $m_F=-3/2$, which gives after
ionization $P\approx-1$.

 For  WFT including 1--3 states,
$\Psi_1\longrightarrow\Psi_3$, and 2--6 states
$\Psi_2\longrightarrow\Psi_6$:
\begin{equation}
B_z(t)=B_0+\frac{dB_z}{dx}vt,\;B_x=B_1(t)\sin\omega t. \label{eq7}
\end{equation}

A tapered electromagnet produces a static magnetic field $B_z(x)$
perpendicular to the beam path  with a field gradient $dB_z/dx$
along the magnet ($x=vt$). Atoms pass through a resonant
radio-frequency (RF) field in the magnetic field which is a slowly
changing function of time because the atoms move through a magnetic
field gradient.

We accepted some parameters the same as in the paper by S.Oh:
$B_0=1.17\times10^{-3}$ T, $dB_z/dx=-1.4\times10^{-2}$ T/m for a
negative static field gradient (or $B_0=4.7\times10^{-4}$ T,
$dB_z/dx=1.4\times10^{-2}$ T/m for a positive  gradient),

$l=5\times10^{-2}$ m, $vt_{final}=5\times10^{-2}$ m,

$\omega=9.63\times10^7$ rad/s for $2\rightarrow6$ transition and
$\omega=1.93\times10^8$ rad/s for $1\rightarrow3$ transition.

It was accepted $v=1.2\times10^3$ m/sec.  The rf amplitude $B_1$ is
of square dependence of $x=vt$ with a zero value at $x=0$ and $x=l$.
$B_1^{max}$ was some units of $10^{-4}$ T.

\begin{figure}[h]
 \centerline{
 \includegraphics[width=50mm,height=50mm]{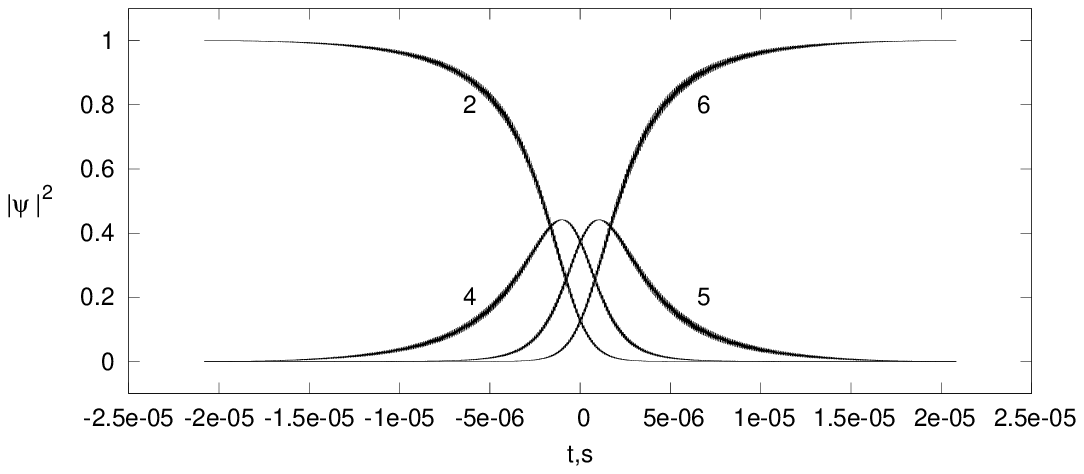}}
 \caption{The probability for transition $2\rightarrow6$}
 \label{Fi2}
\end{figure}

\begin{figure}[h]
 \centerline{
 \includegraphics[width=50mm,height=50mm]{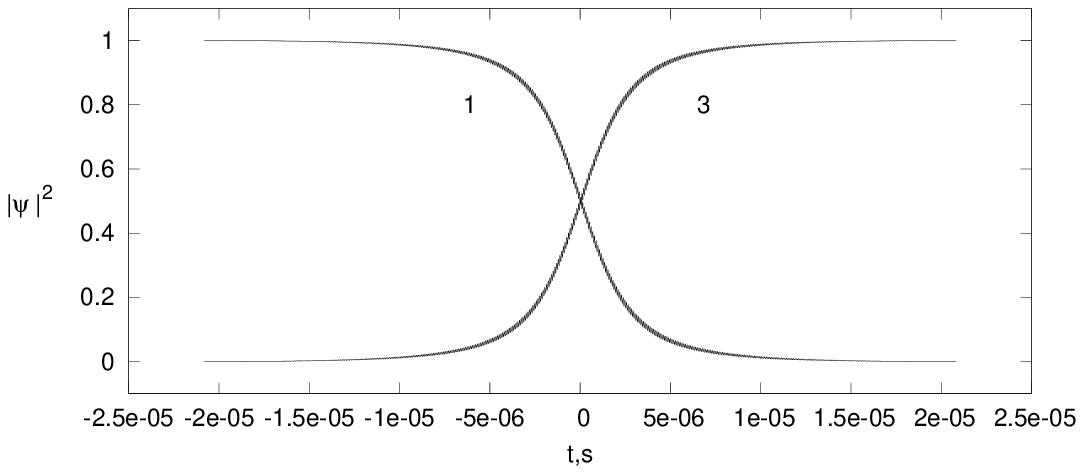}}
 \caption{The probability for transition $1\rightarrow3$}
 \label{F1}
\end{figure}

Some results of computer calculations  in the basis of uncoupled
states for an atom velocity 1200 m/s are presented in the Figs.
\ref{Fi2}, \ref{F1}.

An evident way for producing  polarized helions is to follow the way
of the Laval University group and inject the nuclear polarized
$^3$He$^+$ ions into the electron beam ion source for subsequent
ionization to helions.

 But for a pulsed regime there is a
possibility  of  ionizing metastable atoms directly to $^3$He$^{++}$
and accumulating them in the ion trap of the EBIS  with subsequent
8-$\mu$s-pulse extraction. The ion trap is produced by space charge
of oscillating electrons in a drift tube region (radial confinement)
and  potential barriers on the boundary drift tubes (axial
confinement). The electron cloud is confined  by a solenoid magnetic
field up to 5~T. This high field is also needed to exclude $^3$He
nuclear depolarization in the ionizer.

For  metastable atoms, the cross section of ionization at an
electron energy of 10 keV is
$\sigma^*_i(0\rightarrow1)\approx7.3\times10^{-18}$ cm$^2$, that is,
much larger than for atoms in the ground state
($\sigma_i(0\rightarrow1)\approx2\times10^{-18}$ cm$^2$).

The electron beam density for the ionization to $^3$He$^+$ of
metastable atoms with a velocity of $1.2\times 10^5$ cm/s  at the
length of the ionizer 100 cm  is 30 A/cm$^2$. The cross section
$\sigma_i(1\rightarrow2)$ is $\approx4.3\times10^{-19}$ cm$^2$.
Then, at the electron density 30 A/cm$^2$ the confinement time for
ionization to helions will be 15 ms. With the electron beam diameter
of 5 mm, the effective electron current is 5 A.

The given mode has a number of advantages:

1. It is possible to inject $^3$He into the trap  at the elevated
pressure, which will lead to an increase in the ion charge  in the
trap.

2. Earlier, pulsed extraction of ions from the trap was carried out
for 7 $\mu$s with a current of 1 mA, which corresponded to
$4\times10^{10}$ charges \cite{Donets}. With helium, it is really
possible to get the 2--3 times higher intensity.

The design parameters of the ionizer are as follows:

$\bullet$ electron energy 10~keV,

$\bullet$ effective current $\approx$ 5~A,

$\bullet$ ion trap length 1~m,

$\bullet$ helion  intensity $\approx2\times10^{11}$~ions/pulse.

\begin{figure}[h]
 \centerline{
 \includegraphics[width=50mm,height=50mm]{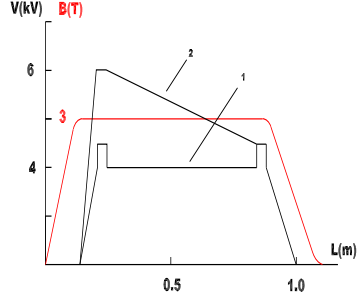}}
 \caption{ The distributions of magnetic field and  voltage  on the
 drift tubes at
accumulation (1) and extraction(2) of the helions.}
 \label{Raspred}
\end{figure}

The experiments \cite{Fimushkin} with the ionizer of the polarized
deuteron source POLARIS (Fig.~\ref{Raspred}) show the feasibility of
storing up to $4\times10^{11}$ charges in the ion trap.

The main elements of the future ionizer are a 5~T superconducting
magnet, electronic optical system, 10~keV modulator, 8~$\mu$s system
of  fast extraction, and remote control system.

At the exit of the ionizer, it is necessary to install a deflecting
magnet for separating $^3$He$^{++}$ and $^3$He$^+$ ions. For
helions, the $G$-factor is $-4.183963$; thus, to get transversal
spin polarization, we need to deflect the  beam by 21.5$^{\circ}$.
Then, a special solenoid should rotate spin to the vertical
direction, up or down.

\section{Depolarization effects}

The time between metastability exchange collisions is $\tau=1/\sigma
v N$, where $\sigma$ is the cross section for the metastability
exchange, $v$ is the velocity of  metastables, and $N$ is the
density of  ground state atoms.

With $v=1.2\times10^5$~cm/s and $\sigma=4\times10^{-16}$~cm$^2$
\cite{Colegrove}, the condition for $\tau\gg T_{acc}$, where
$T_{acc}$ is the time of accumulation, is
$$
\sigma v N\ll T_{acc}^{-1}\,\,\,\,{\rm or}\,\,\, N\ll
2\times10^{10}T_{acc}^{-1},
$$
or $p\ll 6\times10^{-7}/T_{acc}$~Torr.

For $T_{acc}=14$~ms, the required value is $p\ll
4\times10^{-5}$~Torr.

A dangerous process is the symmetric resonant charge transfer
$$
^3{\rm He}^{++}+^3{\rm He}\rightarrow^3{\rm He}+^3{\rm He}^{++}.
$$
The cross section of this process is estimated to be
$\simeq7\times10^{-16}$ cm$^2$ \cite{Schrey}, even larger than the
cross section for metastability exchange.

 Then, it is demanded that
the background pressure of $^3$He be $p\ll10^{-5}\;{\rm Torr}$.

Let a metastable flux be $6\times10^{15}$ atoms/s sterad. If we
assume that the flux of atoms in the ground state is $\simeq10^2$
times higher than the metastable flux, the pressure of the ground
state atoms in the ionizer at the distance of 120~cm from the nozzle
is $\simeq10^{-8}$~Torr, which is acceptable.

When an atom or ion has an electron spin, the nuclear depolarization
can be influenced by  hyperfine interaction. Under an external field
$B$, the primary nuclear polarization of the $^3$He atom is reduced
in the intermediate $^3$He$^{+}$ ion by a factor $\alpha$:
$$
\alpha=1-\frac{1}{2(1+y^2)},
$$
where $y=B/B_c$, with $B_c=0.3087$~T for $^3$He$^{+}$ ions. For
$B=1$~T, $\alpha=0.956$.

The experience of the SATURNE group \cite{Courtois} shows that
depolarization processes seem to be unimportant. They ionized
$^6$Li$^+$ polarized ions to bare nuclei $^6$Li$^{3+}$ in the EBIS
at the field of 5~T without depolarization during accumulation for
3~ms and extraction.

The degree of polarization in the course of acceleration can change
when the spin frequency becomes equal to the integer combination of
the frequencies of betatron and synchrotron motion in the region of
the spin resonance, $\nu=\gamma G=\nu_k$,
\begin{equation*}
\nu_k= k+k_z\,\nu_z+k_x\,\nu_x+k_\gamma\,\nu_\gamma \,,
\end{equation*}
where $\nu_x$ and $\nu_z$ are betatron frequencies, $\nu_\gamma$ is
frequency of synchrotron motion.

Table \ref{t:reson}\cite{Vokal} shows the number of linear
resonances for different particle beams in the Nuclotron ($k$ and
$m$ - integer, $p=8$ - number of superperiods).
\begin{table}[htbp]
\begin{center}
\begin{tabular}{|@{\ \ }l|l|c|c|c|c|}
\hline \hline Resonance type& Resonance condition&
\multicolumn{4}{|c|}{$\vphantom{{d_y^(d)}^(d)}$Number of resonances}\\
\cline{3-6} &
$\vphantom{{d_y^(d)}^(d)}$ & ${}^1H$&${}^2H$&${}^3H$&${}^3He$\\
\hline \hline Intrinsic resonances&
$\vphantom{{d_y^(d)}^(d)}$ $\nu=k\,p\pm\nu_z$& 6&---&8&9 \\
\hline Integer resonances&
$\vphantom{{d_y^(d)}^(d)}$ $\nu=k$ & 25&1&32&37 \\
\hline Non-superperiodical resonances&
$\vphantom{{d_y^(d)}^(d)}$ $\nu=k\pm\nu_z\,(k\neq m\,p)$& 44&2&55&64 \\
\hline Coupling resonances &
$\vphantom{{d_y^(d)}^(d)}$ $\nu=k\pm\nu_x$ & 49&2&63&73 \\
\hline\hline
\end{tabular}
\caption{Linear resonances in the  Nuclotron ring.} \label{t:reson}
\end{center}
\end{table}

\section{Conclusion}

A possibility of developing a polarized helion source for JINR
Accelerator Complex was discussed.

 It seems possible to provide a polarized
beam with polarization larger than 80\% and helion intensity
$\approx2\times10^{11}$~ions/pulse of 8~$\mu$s.

 The depolarizing
effects in the polarized ion source are expected to be low.

 For
acceleration at the NUCLOTRON-M, it is necessary to provide
conditions for low depolarization.

 Installation of the polarized
helion source  at the JINR Accelerator Complex would allow the
program of spin physics experiments to be extended.

\end{document}